\documentstyle[12pt,epsf]{article}
\begin{document}

\newcommand{\sheptitle}
{\Large Intermediate Scale Supersymmetric Inflation, Matter and Dark Energy}
   
\newcommand{\shepauthor}
{G. L. Kane$^1$ and S. F. King$^2$ }

\newcommand{\shepaddress}
{$^1$Randall Physics Laboratory, University of Michigan,
Ann Arbor, MI 48109-1120\\
$^2$Department of Physics and Astronomy,
University of Southampton, Southampton, SO17 1BJ, U.K.}

\newcommand{\shepabstract}
{We consider supersymmetric inflation models in which inflation
occurs at an intermediate scale and which provide a
solution to the $\mu$ problem and the strong CP problem.
Such models are particularly attractive since
inflation, baryogenesis and the
relic abundance of cold dark matter are all related by a set
of parameters which also affect particle physics collider
phenomena, neutrino masses and the strong CP problem.  
For such models the natural situation is a universe containing
matter composed of baryons, massive neutrinos, lightest superpartner cold
dark matter, and axions. The present day relic abundances 
of these different forms of matter are (in principle) calculable
from the supersymmetric inflation model together with
a measurement of the CMB temperature and the Hubble constant.
From these relic abundances one can deduce the amount of the present day
dark energy density.}

\begin{titlepage}
\begin{flushright}
hep-ph/0109118\\
MCTP-01-42\\
SHEP 01-25
\end{flushright}
\vspace{.1in}
\begin{center}
{\large{\bf \sheptitle}}
\bigskip \\ \shepauthor \\ \mbox{} \\ {\it \shepaddress} \\ \vspace{.5in}
{\bf Abstract} \bigskip \end{center} \setcounter{page}{0}
\shepabstract
\begin{flushleft}
\today
\end{flushleft}
\end{titlepage}

\section{Introduction}
Recent data on the cosmic microwave background (CMB) radiation
\cite{cmb} provides strong support for inflation, by measuring
the first, second and third peaks of the angular power spectrum.
Inflation therefore seems to be increasingly well established.
Low energy supersymmetry (SUSY) is perhaps less well established 
(though there is considerable indirect evidence for it \cite{gordy0}), 
but is certainly 
desirable from many points of view, and has the advantage
when combined with inflation of helping to ensure that 
the inflaton potential is sufficiently and naturally flat \cite{toni}.
The CMB data also supports a $\Lambda CDM$ universe in
which the energy density is dominated by dark energy (DE)
(corresponding possibly to a cosmological constant $\Lambda$), 
and cold dark matter (CDM) \cite{lcdm}.

In this paper we consider supersymmetric inflation 
models in which inflation occurs at an intermediate scale
\cite{kane,Bastero-Gil:1998vn,leptogenesis,sugra},
and which provide an
intermediate scale solution to the $\mu$ problem and
a solution to the strong CP problem via the 
Peccei-Quinn (PQ) mechanism \cite{PQ}.
When right-handed neutrinos are included, such models may give
baryogenesis via leptogenesis. They will have 
calculable CDM and DE relic densities.
And they may have supergravity
(SUGRA) mediated SUSY breaking with no moduli or gravitino problems.
A main point of the paper is to show that in such models
in which all these features are simultaneously present
there will be fewer parameters than in models in which these
problems are separately addressed. We will discuss an explicit
example of such a model in order to demonstrate the
connections between the physics of each of the 
separate sectors, and the resulting enhanced predictivity leading for example
to connections between collider physics experiments and cosmological
observations. 

Within such a framework one may expect on general grounds
that our Universe contains sizeable relic abundances of baryons
(from leptogenesis), axions (a) (from the solution to the strong CP
problem), as well as weakly interacting massive particles
(WIMPS). In R-parity conserving SUSY the WIMP is identified
as the lightest supersymmetric particle (LSP), and this is
often assumed to be
the lightest neutralino of the minimal supersymmetric standard model
(MSSM). We shall argue from this perspective 
that the LSP could equally well be a lighter stable singlet
(singlino) \cite{singlino}
identified with an axino \cite{axino} or inflatino \cite{inflatino}.
Neutrino masses provide smaller amounts of hot dark matter.
Within this approach all these forms
of matter in the universe will have calculable
relic abundances, given
a measurement of the CMB temperature and the Hubble constant,
which are related to the parameters 
of the underlying supersymmetric theory which may be
determined from particle physics experiments.
This amounts to a generalisation of the observation
made a few years ago that the relic abundance of
neutralinos is related to the parameters of the supersymmetric
theory. Further, a given parameter
typically is relevant to more than one relic abundance, so the total
number of parameters is fewer than when the various forms are considered
separately. From the calculated present day relic abundances of matter 
one can deduce the amount of the present day
DE density, even without specifying the physics of DE.

\section{The Omega Problem in Supersymmetric Inflation}
In general the ratio of the total density of the universe $\rho_{tot}$
to critical density $\rho_{crit}$
is given by $\Omega_{tot}$ where
\begin{equation}
\Omega_{tot} = \Omega_{\gamma}+\Omega_{matter}+ \Omega_{DE} 
\label{omega}
\end{equation}
and $\Omega_{\gamma}$, $\Omega_{matter}$, $\Omega_{DE}$ are the ratios of
radiation density $\rho_{\gamma}$, 
matter density $\rho_{matter}$
and DE density $\rho_{DE}$ to critical
density $\rho_{crit}$, and the radiation density is unimportant 
$\rho_{\gamma}\ll \rho_{matter}$.
Note that the critical density is a function of time 
and in the present epoch
$\rho_{crit}=3M_P^2H_0^2 =(3h^{1/2}\times 10^{-3}\ {\rm eV})^4\sim
(M_W^2/M_P)^4$ where $M_W$ is the weak scale, $M_P$ is the Planck
scale, and $H_0=100h {\rm km.s^{-1}Mpc^{-1}}$ 
is the present day Hubble constant, with $h=0.7\pm 10\%$
\cite{freedman}.
From observation \cite{cmb} $\Omega_{DE}\sim 2/3$
while $\Omega_{matter}\sim 1/3$ and $\Omega_{tot}$ is very close to unity.
Inflation predicts $\Omega_{tot} = 1$, for all times after
inflation.
The matter contributions consist of (at least)
\begin{equation}
\Omega_{matter}=\Omega_{b}+\Omega_{\nu}+\Omega_{LSP} + \Omega_{a}  
\end{equation}
The most recent data \cite{cmb} is consistent with nucleosynthesis
estimates of $\Omega_{b}\sim 0.04$, where the baryons (b) in the universe
are mainly to be found in dark objects. 
The determined value of 
$\Omega_{CDM}\sim 0.3$ contains unknown relative contributions from
$\Omega_{LSP}$ and $\Omega_{a}$. 
Super-Kamiokande sets a lower limit on neutrino masses
$\sum_i m_{\nu_i}\geq 0.05$ eV which corresponds to $\Omega_{\nu}\geq 0.003$. 
In hierarchical neutrino mass models the lower bound is saturated
and interestingly the neutrino density is then comparable to
the visible baryon density $\Omega_{stars}\sim 0.005$.

We now discuss how to calculate the relic densities from our present
rather primitive state of knowledge about the comprehensive theory needed
to really do that. \footnote{The calculation 
will involve non-perturbative cosmological effects 
during the reheating process, such as preheating
\cite{preheating1,preheating2,preheating3,preheatingus}.}
In the following it is important to keep in mind that our
present inability to calculate the ratios of different forms of matter
should be distinguished from the ability to calculate them in principle.
Later we discuss a simple model which illustrates these ideas.

Of the three likely dominant densities
$\Omega_{b}$, $\Omega_{LSP}$ and $\Omega_{a}$,
the latter arises from non-relativistic axions being produced 
at the QCD scale by the usual misalignment mechanism, and is the most
difficult of the three to estimate since it depends on a randomly
selected angle. \footnote{Any relativistic axions
produced during (p)reheating will be red-shifted away.}
Therefore we shall focus mainly on the question
of how to calculate $\Omega_{b}$ and $\Omega_{LSP}$.

The present day value of $\Omega_{b}$ corresponds to
the ratio of baryon number density to entropy
density of the universe 
$Y_b=n_b/s\approx 0.7\times 10^{-10}$, assuming $n_{\bar{b}}=0$.
For the calculation of $Y_b$, 
we shall restrict ourselves here to the so-called
leptogenesis mechanism. 
The basic idea of leptogenesis \cite{yanagida1,luty}
is that
right-handed neutrinos (or possibly sneutrinos) 
are copiously produced in the early universe,
then decay to produce lepton number (and hence $B-L$) asymmetry. 
The lepton number asymmetry is subsequently converted
into baryon number asymmetry by sphaleron interactions.
Concerning the three Sakharov conditions: 
the CP-violation originates from complex Yukawa
coupling constants \footnote{Complex SUSY mass parameters are not relevant for
leptogenesis if its scale is above that of SUSY breaking.}; 
lepton number violation originates from
the Majorana mass of the right-handed neutrinos,
and baryon number violation from sphalerons.
Concerning the out-of-equilibrium condition,
in the conventional approach to leptogenesis it is assumed that
the right-handed neutrinos are produced by their couplings
to other particles in the thermal bath, but that these
couplings are sufficiently weak that the decays occur
out-of-equilibrium, leading to a narrow range of couplings
\cite{Buchmuller:2000as,Hirsch:2001dg}. 
From the perspective of inflation the conventional leptogenesis picture
will change if the reheat temperature is below the mass of the 
lightest right-handed neutrino. In this case right-handed neutrinos 
may be produced during reheating via 
(direct or indirect) couplings to the inflaton field,
and may then be produced
with masses greatly exceeding the reheat temperature, 
providing only that they are lighter than the inflaton field.
In this case the out-of-equilibrium condition is automatically satisfied
during reheating. This second mechanism is preferred from the point
of view of the gravitino constraint, since in this case if
the reheat temperature is 
below the limit $T_R<10^9$ GeV then thermally produced 
gravitinos are not a problem.\footnote{If the gravitino is the LSP
then this limit on the reheat temperature may be relaxed
\cite{Bolz:1998ek}.}

In the conventional case the LSP is regarded as the lightest
neutralino $\tilde{\chi}_1$ 
(a linear combination of neutral bino $\tilde{B}$,
wino $\tilde{W}_3$, and higgsinos $\tilde{H}_1$, $\tilde{H}_2$)
and it is produced in thermal equilibrium at 
temperatures above its mass $\sim M_W$. 
According to the following very crude argument
when their annihilation rate
$\Gamma \sim \sigma n_f  \sim n_f/M_W^2$ becomes smaller
than the expansion rate of the universe 
$H\sim T_f^2/M_P$ (at a temperature $T_f\sim M_W$), they freeze out of the
thermal bath, and the present day matter density is then
$\rho_{\chi} \sim M_Wn_f(T/M_W)^3\sim M_W^2T^3/M_P$
including the dilution factor 
$(R_f/R)^3\sim (T/M_W)^3$.
The current temperature is obtained by roughly equating
$\rho_{\chi}\sim \rho_{\gamma}\approx T^4$ which gives
$T\sim M_W^2/M_P$. Inserting this temperature we find
$\rho_{\chi}\sim (M_W^2/M_P)^4 \sim \rho_{crit}$.
Many more careful analyses have been performed in order to obtain
precise estimates for $\rho_{\chi}$ by considering the detailed
annihilation channels within different regions of SUSY parameter
space. What concerns us more here is how including the
effects of inflation will change this simple framework.
As in the discussion of leptogenesis (above) 
from the perspective of inflation 
and baryogenesis, and in general a broader picture,
the conventional picture
will change if the reheat temperature is below the mass of the 
lightest neutralino, and they are produced by non-thermal
processes during the reheating process after inflation.
Another way in which the physics might differ is if the
LSP is not the lightest neutralino, but instead some lighter singlino
associated with an axino or inflatino. For example, if
the lightest neutralino freezes out then decays into an axino ($\tilde{a}$)
then the present day axino density is suppressed 
by the ratio of axino mass to lightest neutralino mass
$\rho_{\tilde{a}}=(m_{\tilde{a}}/m_{\chi})\rho_{\chi}$ \cite{axino}.
In order to learn the actual
relic density of LSPs (or axions or any candidate) it is necessary to
actually calculate it; detecting LSPs or axions is possible even if the
relic density is well below $\Omega_{matter}$ of order 1/3 \cite{gordy}.

\section{An Example}
\subsection{Why Intermediate Scale Inflation?}
We now wish to consider a specific example of a
model which addresses the particle
physics issues mentioned earlier, in order to illustrate many of the
general features that we have discussed above.
The brand of inflation most closely related to particle physics
seems to be hybrid inflation which may occur at a scale
well below the Planck scale, and hence be in the realm of particle
physics \cite{Linde:1994cn}. The next question is what is the
relevant scale at which hybrid inflation takes place?
One obvious possibility is to
associate the scale of inflation with some grand unified theory
(GUT) symmetry breaking scale, as originally conceived by Guth
\cite{Guth:2000ka}. However it is somewhat ironic that 
hybrid inflation at the GUT scale faces the magnetic monopole
problem, which was precisely one of the original motivations
for considering inflation in the first place! Although in certain
cases this problem may be resolved, there are typically further symmetry
breaking scales below the GUT scale at which discrete symmetries
are broken, leading to problems with cosmological domain walls.

As an example of the difficulties faced by GUT scale inflation models,
consider the breaking of the Pati-Salam symmetry
group $SU(4)\times SU(2) \times SU(2)$ down to the standard
model gauge group. The minimal symmetry breaking potential
in this model \cite{King:1998ia}
contains a singlet which could be a candidate
for the inflaton of hybrid inflation.
However it was immediately realised that 
such a scenario would face a magnetic monopole problem
since the gauge group is unbroken
during inflation, and only broken by the choice of vacuum after
inflation \cite{King:1998ia}.
A possible solution to this problem is to consider the
effect of higher dimension operators in the superpotential which
for a range of parameters
have the effect of breaking the gauge symmetry during inflation
\cite{Jeannerot:2000sv}. Such a scheme can also address the 
$\mu$ problem and the strong CP problem, at the expense of introducing
additional singlets which develop vacuum expectation values
(VEVs) at the PQ symmetry breaking
scale \cite{PQ}. 
However in this scheme it turns out that the vacuum in which 
PQ symmetry is unbroken is preferred below the reheat temperature 
of the model so that PQ symmetry is not broken after inflation
\cite{Jeannerot:2000sv}. Moreover PQ symmetry breaking is associated
with the breaking of discrete symmetries in the model, so that it
would lead to the domain wall problem in any case
\cite{Jeannerot:2000sv}. The only solution to these problems
seems to be to assume that PQ symmetry is also broken during
inflation, but since the inflaton has zero PQ charge, and since 
the inflation scale of order $10^{14}{\rm GeV}$
\cite{Jeannerot:2000sv} and hence much larger
than the PQ symmetry breaking scale, this assumption seems rather
questionable. 

Intermediate scale hybrid inflation 
immediately solves
both the magnetic monopole problem and the domain wall problem.
The idea is simply that there is a period of hybrid inflation
occuring below the GUT scale at the PQ symmetry breaking scale
itself, in which the inflaton carries PQ charge and so the choice of
domain is fixed during inflation. The universe therefore inflates
inside a particular domain, and the magnetic monople relics
produced by the GUT scale symmetry breaking are inflated away.
This provides a powerful
motivation for intermediate scale inflation, which is
is the subject of this paper. We now turn to an explicit
example of an intermediate scale inflation model.

\subsection{Intermediate Scale Supersymmetric Inflation Model}
The model we consider \cite{Bastero-Gil:1998vn} is a variant of the 
NMSSM. This model has a SUGRA foundation \cite{sugra},
and leptogenesis and reheating has been studied \cite{leptogenesis},
and preheating \cite{preheatingus,gravitino} has been demonstrated
not to lead to over-production of either axions or gravitinos.
The model provides a solution to the strong CP problem and the
$\mu$ problem, with inflation directly solving the monopole
and domain wall problems at the inflation scale.
It is therefore a well motivated, successful model 
that has been well studied and does not appear to suffer
from any embarrassing problems, and is therefore a suitable
laboratory for our discussion here.
This variant of the NMSSM has the following superpotential terms involving the
standard Higgs doublets $H_{u},H_{d}$
and two gauge singlet fields $\phi$ (inflaton) 
and $N$,
\begin{equation}
 W = \lambda N H_{u} H_{d} - k \phi N^{2}  \label{eq:super}
\end{equation}
where $\lambda ,k$ are dimensionless coupling constants.
Notice that the standard NMSSM is recovered if we replace the inflaton $\phi$ 
by N.  However this leads to the familiar domain wall problems arising from 
the discrete $Z_{3}$ symmetry.  In this new variant, the $Z_{3}$
becomes a global Peccei-Quinn $U(1)_{PQ}$ symmetry that is commonly 
invoked to solve the strong CP problem \cite{PQ}.  
This symmetry is broken in the true 
vacuum by intermediate scale $\phi$ and N VEVs, 
where the axion is the pseudo-Goldstone
boson from the spontaneous symmetry breaking and constrains the size of the 
VEVs. With such large VEVs
this model should be regarded as giving an intermediate scale
solution to the $\mu$ problem, and as such will have the
collider signatures discussed in \cite{singlino}.

We can make the $\phi$-field real by a choice of the (approximately) massless 
axion field.  We will now regard $\phi$ and $N$ to be the real components of 
the complex singlets in what follows.
When we include soft SUSY breaking mass terms,
trilinear terms $A_{k}k \phi N^{2} + h.c.$ (for real $A_{k}$) and neglect the
$\lambda N H_{u} H_{d}$ superpotential term, we have the following 
potential:
\begin{equation}
 V=V_{0} + k^{2} N^{4} + \frac{1}{2} m^{2}(\phi) N^{2} 
  + \frac{1}{2} m_{\phi}^{2} \phi^{2}
     \label{eq:potential} 
\end{equation}
where $m^{2}(\phi)= m_{N}^{2} + 4k^{2} \phi^{2} - 2k A_{k} \phi $.
We can identify the various elements of the potential: $V_{0}$ arises
from some other sector of the theory, SUGRA for example, and dominates the
potential \cite{sugra}; 
the soft SUSY breaking parameters $A_{k}$ and $m_{N}$ are
generated through some
gravity-mediated mechanism with a generic value of $O(TeV)$.

The basic idea of hybrid inflation is very simple \cite{toni}.
For the field dependent mass of the $N$ scalar
positive, $m^2(\phi)>0$, then $N=0$ since its potential has
positive curvature. With $N=0$ the potential becomes
very simple indeed,
\begin{equation}
V=V_0+\frac{1}{2}m_{\phi}^2\phi^2
\label{simple}
\end{equation}
We shall assume that $m_{\phi}$ comes from no-scale SUGRA, and
vanishes at the Planck scale \cite{sugra}, so that
it is generated through radiative
corrections such that $m_{\phi}^{2} \sim -k^{2} A_{k}^{2}$. 
Since $m_{\phi}^{2}$ is negative, during inflation
$\phi$ is slowly rolling away from the origin,
and therefore we have {\it inverted} hybrid inflation.
When $\phi$ exceeds a critical value
\begin{equation}
\phi_c= ({A_{k}}/{4k})(1 - \sqrt{ 1 -{4 m_{N}^{2}}/{A_{k}^{2}}})
\end{equation}
the sign of the field dependent mass of 
$N$ will become negative, $m^2(\phi)<0$, and the $N$ field
will no longer be pinned to zero, but will roll out to the
global minimum of the potential, 
\begin{equation}
 \langle \phi \rangle = {A_{k}}/{4k}, \ \ 
 \langle N \rangle = ({A_{k}}/{2\sqrt{2} k}) 
\sqrt{ 1- {4 m_{N}^{2}}/{A_{k}^{2}}}.
\end{equation}
Together with our assumptions about the SUSY parameters, this implies
\begin{eqnarray}
A_{k} \sim k \phi_{c} \sim k \langle N \rangle 
\sim k \langle \phi \rangle \sim
\lambda \langle N \rangle \equiv \mu \sim 10^{3} \, GeV
\label{parameters}
\end{eqnarray}
During inflation the inflaton field $\phi$ must
satisfy the slow roll conditions $\epsilon,\eta \ll 1.$ 
\footnote{Recall that
$\epsilon \equiv \frac{1}{2} \tilde{M}_{P}^{2} 
(V'/V)^{2}$ 
with $| \eta | \equiv | \tilde{M}_{P}^{2} V''/V |$, 
where $V' (V'')$ are the first (second) derivatives of the potential
and $\tilde{M}^{2}_{P} = M^{2}_{P}/8 \pi$ is the reduced Planck mass.
The COBE normalisation is given by
$ \delta_{H}^{2} = ({1}/{150 \pi^2}){V_0}/({\tilde{M}_{P}^{4} \epsilon})$.}
The COBE normalisation $ \delta_{H} =1.95 \times 10^{-5}$ requires
the value of the inflaton mass from radiative corrections
$m_{\phi}^{2} \sim -k^{2} A_{k}^{2} \sim -(100eV)^{2}$
and from Eq.\ref{parameters}
this implies that $\lambda ,k \sim 10^{-10}$ and 
$\langle N \rangle \sim \langle \phi \rangle \sim 10^{13}$
GeV. We address the smallness of $\lambda, k$ in the next section.
The spectral index $n$ which relates to the 
power spectrum $P_{k} \propto k^{n-1}$, is given by
$|n-1| = 2\eta - 6 \epsilon \sim 2\eta \sim 10^{-12}$ which provides a
basic prediction of the model. The present value of the spectral index
in the range $0.80<n<1.05$ at 68\% C.L. slightly disfavours the accurate
prediction that $n=1.00$ but only at the 1$\sigma$ level,
and we may have to wait for the results from the Planck satellite
which will measure it to an accuracy of $\Delta n=\pm 0.01$
to definitively test this prediction \cite{liddle}.

It is non-trivial that a set of parameters exists that is
consistent with axion and SUSY physics and allows
the correct COBE perturbations to be achieved by radiative corrections.
Without SUSY one would be free to add soft scalar masses at will,
but with SUSY one must rely on the theory which either generates
soft masses of order a TeV, or sets them equal to zero as in no-scale
SUGRA, in which case
the radiative corrections, which are under control in the case
of SUSY predict the relevant value of the soft parameters, without
any further adjustable parameters. Thus SUSY is playing a crucial
role in the model which is why we refer to it as a Supersymmetric
Inflation Model.

\subsection{Role of Non-renormalisable Operators}
The couplings $\lambda,k$ should be thought of as effective couplings 
arising from 
non-renormalizable operators \cite{kane,Bastero-Gil:1998vn} so
they are actually couplings of order unity times small factors arising from 
ratios of VEVs to the Planck mass to some power, as can occur in
models, and are not unnaturally small. In the original model
it was suggested that the superpotential in Eq.\ref{eq:super}
arose as an effective theory from a non-renormalisable superpotential
which at leading order is given by
\begin{equation}
 W_{NR} = \lambda' N H_{u} H_{d}\frac{M\overline{M}}{M_P^2}
 - k' \phi N^{2}\frac{\overline{M}^2}{M_P^2}
+c\frac{(M\overline{M})^3}{M_P^3}
+d\frac{(N\overline{M})^5(M\overline{M})^2}{M_P^{11}}+\cdots
\label{eq:super2}
\end{equation}
where two extra singlets $M$ and $\overline{M}$ have been introduced
which develop VEVs by a radiative mechanism
$<M>=<\overline{M}>\sim 10^{14}{\rm GeV}$, as a result of which 
we recover the original superpotential in Eq.\ref{eq:super} and we 
reinterpret our couplings $\lambda, k$ as effective couplings given by
\begin{equation}
\lambda\equiv \lambda'\frac{<M><\overline{M}>}{M_P^2}\sim \lambda'
10^{-10}, \ \ 
k\equiv k'\frac{<\overline{M}>^2}{M_P^2}\sim k'10^{-10},
\end{equation}
Thus the underlying coupling constants $\lambda' ,k'$ are of order
unity, although for the most part we find it convenient to 
discuss the model in terms of effective couplings $\lambda ,k$.
The underlying theory respects a $Z_3\times Z_5$ symmetry,
and the $U(1)_{PQ}$ symmetry arises as an approximate effective
symmetry, leading to an explicit contribution to the axion mass
from the term proportional to $d$ 
which tilts the axion potential slightly, and perturbs the
$\theta$ angle by an amount $\Delta \theta \approx 10^{-11}$,
thereby preserving the PQ solution to the strong CP problem,
but providing a prediction for the next generation of
electric dipole moment (EDM) experiments.

\subsection{The Cosmological Constant Problem}
Notice that the SUGRA-derived potential contribution $V_{0}$ exactly cancels
with the other terms (by tuning) to provide agreement with the observed small
cosmological constant. 
Thus we assume that at the global minimum
$V(\langle \phi \rangle , \langle N \rangle )=0$ which implies that
$ V_0 = k^{2} \langle N \rangle^{4}$.
The height of the potential during inflation is therefore 
$V_{0}^{1/4} = k^{1/2} \langle N \rangle \sim 10^{8} GeV$.
Since the approach has a consistent way to set the large 
cosmological constant to zero, the absence of a real solution to this problem may not
be an obstacle to the implications of the approach.

\subsection{Parameter Counting and Singlino Mixing}
A relevant parameter count at this stage reveals two superpotential
effective parameters ($\lambda$ and $k$),
the two soft SUSY breaking parameters ($A_{k}$ and $m_{N}$),
plus the constant energy density $V_0$.
From these five parameters we have inflated the universe
with the correct COBE perturbations,
provided a $\mu$ term of the correct order of magnitude
and solved the strong CP problem.
They also govern the phenomenology of the singlet Higgs
and Higgsino components of $\phi$ and $N$ which may weakly mix
with the MSSM superfields $H_u,H_d$. For example the
Higgsino mixing matrix in the basis 
$\tilde{H}_d,\tilde{H}_u,\tilde{N},\tilde{\phi}$ is
\begin{equation}
\left( \begin{array}{cccc}
0 & -\lambda<N> & -\lambda<H_u> & 0   \\
-\lambda<N> & 0 & -\lambda<H_d> & 0   \\
-\lambda<H_u> & -\lambda<H_d> & 2k<\phi > & 2k<N>   \\
0 & 0 & 2k<N> & 0
\end{array}
\right),
\label{higgsinos}
\end{equation}
The LSP will be the lightest eigenvalue of the full ``ino'' matrix,
extended in the usual way to include gaugino-higgsino mixing.
Clearly if $k<\lambda/2$ then a singlino will
be the LSP. In our case the singlino may be regarded as
a linear combination of axino and inflatino.
The coupling of the superfield $S$ containing the
singlino $\tilde{S}$ is given from the usual result \cite{singlino}
$W=\mu(1+\epsilon S/v)H_uH_d$ with $\mu = \lambda <N>$.
Here $\epsilon \sim v/f_a$ where the $f_a$ is the axion decay constant 
and $v$ is an electroweak VEV. Thus we have
$f_a \sim <N>$ so $\epsilon \sim \lambda$, and hence
\begin{equation}
W\sim \lambda<N>(1+\lambda S/v)H_uH_d.
\end{equation}
As usual in models based on an intermediate scale solution to the
$\mu$ problem \cite{kane,Bastero-Gil:1998vn,singlino} the coupling of the 
singlino to the neutralinos means that $\tilde{S}$ nearly decouples.
However the conservation of R-parity means that eventually the
lightest neutralino produced in colliders must decay into the
singlino, and all the collider signatures discussed
in \cite{singlino} may apply. In the case that the lightest
neutralino leaves the detector before it decays into the singlino,
there will be no unconventional collider signature. In this case the
knowledge concerning a lighter singlino will
come from cosmology since the LSP relic density gets diluted
by the ratio of the singlino to lightest neutralino masses,
and direct dark matter searches will not see anything since the
singlino LSP will not scatter off nuclei.

\begin{table}[tbp]
\hfil
\begin{tabular}{cccccc}
\hline \hline 
 & $V_0$ & $k$  & $\lambda$ & ${\cal L}_{soft}$ &  ${\cal L}_{Yuk}$          
\\ \hline \hline
Inflation and COBE 
& $\surd$ & $\surd$  & -  &  $\surd$ & - \\
MSSM $\mu$ parameter 
& -      & - &  $\surd$ &  $\surd$ &   - \\
Fermion Masses, Mixings
& - &   - & -  & - & $\surd$ \\
SUSY collider physics
& - & $\surd$ & $\surd$ & $\surd$ & $\surd$ \\
Strong CP, axion abundance ($\Omega_a$)
& -  &    $\surd$ & $\surd$ & $\surd$ & - \\
Leptogenesis ($\Omega_b$)
& - & $\surd$ & $\surd$ & $\surd$ & $\surd$ \\
LSP abundance ($\Omega_{LSP}$)
& - & $\surd$ & $\surd$ & $\surd$ & - \\
\hline \hline
\end{tabular}
\hfil
\caption{\footnotesize 
This table illustrates the fact that a particular parameter of the
model (columns) simultaneously controls
several different aspects of particle physics
and cosmology (rows) which are thereby related.
${\cal L}_{soft}$ contains $A_k,m_N^2$ and the other soft parameters,
${\cal L}_{Yuk}$ contains the Yukawa coupling constants
controlling all fermion masses and mixing angles.}
\label{forbidden}
\end{table}

One of the main things we want to emphasize is
the connection between the calculation
of relic densities and the other physics, via their common
parameters. This is summarised in Table 1 for the particular model
we have been discussing.
The same parameters that control the ino mass matrix
will also be involved in reheating of the universe after inflation, and
giving the relic densities of LSP and in leptogenesis
as we discuss in the next section. 
Different models may have different mechanisms to solve
some of the problems, different reheating and preheating, and so on, but will
still lead to a version of Table 1.

\subsection{Preheating/Reheating}
It is usually assumed that
inflation ends with the singlets $\phi,N$ oscillating about their global 
minimum. Although the final reheating temperature
is estimated to be of order 1 GeV \cite{Bastero-Gil:1998vn},
during the reheating process the effective temperature
of the universe, as determined by the radiation density,
can better be viewed as rapidly rising to
$V_{0}^{1/4} = k^{1/2} \langle N \rangle \sim 10^{8} GeV$
then slowly falling
to the final reheat temperature \cite{leptogenesis}
during the reheating process.
This {\it reheating} gives entropy to the Universe.  
Non-perturbative effects can produce particles with masses up to the potential
height, i.e. $m \leq V_{0}^{1/4} \sim 10^{8} GeV$ ({\it preheating}).
The preheating and reheating process in this model is 
quite complicated, but the essential physics is as follows.
To begin with the potentially problematic axions and gravitinos are not over 
produced during preheating ~\cite{preheatingus,gravitino}.  
Higgs scalars are copiously produced through preheating via the
couplings $\lambda$ and $k$ to the oscillating inflaton fields.
Although the neutralinos are produced in Higgs decays via preheating,
once the Higgses decay they go into thermal equilibrium, and 
subsequently freeze out while the universe is radiation
dominated, similar to the usual hot big bang scenario. 
However, for a range of
parameters the singlino is lighter than the lightest neutralino,
and in this case after freeze-out the lightest neutralino decays into the 
singlino thereby reducing the LSP relic density by the ratio
of their masses. 

Recently it has been realised that reheating in 
all hybrid models, including the SUSY motivated ones of interest to
us here, goes through very effective tachyonic preheating
\cite{preheating3}. As a result the stage of the
background oscillations of the scalars around the minimum
of the potential will never be reached. This picture is dramatically
different from the early papers on the reheating in hybrid models
\cite{preheating1,preheating2} and will probably affect the
results in \cite{preheatingus}. On the other hand the 
new picture of preheating implies that exciting one field 
(for example $\phi$ or $N$) is sufficient to rapidly drag all other
light fields with which it interacts into a similarly excited state.
This strengthens our claim that the Higgs doublets which interact with
$N$ will be efficiently preheated.

Once the Yukawa couplings are included in the superpotential,
right-handed sneutrinos are also expected to be produced 
during the initial period of preheating via their
couplings to the Higgs doublets,
and decay out-of-equilibrium
into leptons and Higgses giving rise to leptogenesis.
We have already remarked that, unlike the usual hot big bang scenario,
the out-of-equilibrium condition is automatically satisfied during
reheating, and furthermore the production mechanism of right-handed
neutrinos is totally different.
In the standard scenario the baryon asymmetry 
is given by $Y_b \sim  d \epsilon_1/g^*$
where $\epsilon_1$ is the lepton number asymmetry produced
in the decay of the lightest right-handed neutrino of mass $M_1$, 
$g^*$ counts the effective 
number of degrees of freedom (for the SM $g^* = 106.75$ while for the 
Supersymmetric SM $g^* = 228.75$) and $d$ is the dilution 
factor which takes into account suppressions from either the couplings
being too small to thermally produce right-handed neutrinos,
or too large to satisfy the out-of-equilibrium condition.
Typically $d\ll 1$ except for a narrow range of couplings
\cite{Buchmuller:2000as}. In the inflationary picture
of reheating outlined above, the baryon asymmetry is
given by $Y_b \sim  \gamma {\epsilon_1}(cV_0)^{1/4}/M_1 $
where $c$ is the fraction of the total vacuum energy converted
into right-handed neutrinos due to preheating,
and $\gamma$ accounts for dilution due to entropy production
during reheating. 
Since $\epsilon_1 \sim 10^{-6} (M_1/10^{10}GeV)$
\cite{Hirsch:2001dg} we find that 
$Y_b \sim 10^{-8}\gamma (c)^{1/4}$,
apparently independently of $M_1$ (although $c$ will depend
on $M_1$, for instance $c=0$ for $M_1>10^8$ GeV.)

Solving the Boltzmann equations for a particular choice of parameters
in this model the densities of the neutralinos, radiation,
relativistic axions and baryons were calculated at reheating time,
defined as the time at which the oscillating singlet energy density
rapidly decayed to zero \cite{leptogenesis}.
This time represents the start of the hot big bang.
The important point to emphasise is that at this time $t_{RH}$,
for a given model the Boltzmann equations enable us to calculate
the energy density of the
different types of matter $\rho_{matter}(t_{RH})$, as well
as the radiation energy density $\rho_{\gamma}(t_{RH})$.
In the present model the details of this calculation are discussed
in \cite{leptogenesis}, and a similar calculation may be performed
for any other model.

\section{How To Calculate the Size of the 
Dark Energy Density in Supersymmetric Inflation}
We now turn to the question of dark energy, which is not addressed by
our supersymmetric particle physics based model of inflation.
The origin of DE might be a traditional cosmological constant
with equation of state $w=-1$ or some time-varying smooth energy
(quintessence) with $-1<w<-0.6$ where the upper bound is from
current observations, and may be in conflict with some quintessence
models \cite{krauss}.
\footnote{Note that for a scalar field
$w=p/\rho=(KE-PE)/(KE+PE)$ and if the kinetic energy (KE)
is small compared to the potential energy (PE) then $w$ is negative. 
$w<-1$ would require negative KE which corresponds to the scalar
being a ghost field, and a loss of unitarity \cite{cline}.}
Quintessence models assume a zero cosmological constant and add
some arbitrary field to account for DE.
From the point of view of our inflation model, the simplest
possibility is to assume that at the global minimum (after inflation) 
the height of the potential is not zero but about 
$10^{-3}{\rm eV}$. This possibility, which just corresponds to
a standard cosmological constant with $w=-1$, 
can be arranged (though not explained) by tuning $V_0$
in Eq.\ref{eq:potential}; $V_0$ has to be tuned in any case to 
give a zero cosmological constant, so this possibility requires
no additional tuning to that already required in the model.
In the no-scale SUGRA model \cite{sugra} $V_0$ arises from
the moduli fields in the string theory, and is determined by the
non-perturbative physics of moduli stabilisation which is not yet understood
from a fundamental point of view but may nevertheless be parametrised.
Of course such a proceedure raises the cosmic coincidence
question: why should we have $\rho_{DE}\sim \rho_{matter}$ at the
present epoch? Until the cosmological constant problem is resolved,
there is no way that this question can be answered.
We reject recent claims to the contrary which are based on setting the
cosmological constant to zero by hand to start with, since there is
always the danger in this approach that one has thrown away the
baby along with the bathwater.
In the absence of anything better,
some authors have turned to anthropic arguments, but many
anthropoids reject this approach also as long
as alternatives may be possible. 

Is there anything that we can say about DE at the current time
from the perspective
of our supersymmetric particle physics based model of inflation?
Perhaps surprisingly the answer is positive: we shall show that 
we can deduce the present day value of dark energy density from the
model, together with the measured CMB temperature
and the Hubble constant, even if the model
does not yet specify the physics of dark energy!

A key point of our approach is that a supersymmetric particle physics based
model of inflation enables us to calculate (in principle at least)
the (energy or number)
densities of all forms of radiation and matter 
(but excluding dark energy) at some
early time $t_{RH}$ after inflation and reheating has taken place,
corresponding to the start of the standard hot big bang.
For simplicity we consider only
one type of matter energy density $\rho_{matter}(t_{RH})$
(which may readily be obtained from the calculated number density) 
and radiation energy density $\rho_{\gamma}(t_{RH})$.
The argument may be
straightforwardly generalised to the case of several components
of radiation and matter.
Now, using the equations of the standard hot big bang,
we wish to obtain their values at the present time $t_0$, 
$\rho_{\gamma}(t_{0})$ and $\rho_{matter}(t_{0})$.
Without further information this is impossible since we need something to 
tell us when the present time $t_0$ is, and moreover the model does
not specify either $\rho_{DE}(t_{RH})$, or its equation of state,
both of which will influence the evolution of the universe.
Therefore let us input into our analysis the present day
observed CMB temperature $T_0=2.725{\rm K}$, which corresponds to
a photon density $\rho_{\gamma}(t_{0})=(2.115\times 10^{-4}{\rm eV})^4$,  
a photon number density $n_{\gamma}=410 {\rm cm^{-3}}$,
and, assuming three families of light neutrinos, an entropy density
$s=7.04n_{\gamma}$. 
Then, ignoring additional sources of entropy
between $t_{RH}$ and $t_0$ (such as electron-positron annihilation),
since we know the equations of state for photons and matter,
$\rho_{\gamma}\sim R^{-4}$ and $\rho_{matter}\sim R^{-3}$,
where $R$ is the scale factor of the universe, using the initial
values of $\rho_{\gamma}(t_{RH})$ and $\rho_{matter}(t_{RH})$
predicted by the model and the present value of 
$\rho_{\gamma}(t_{0})$ from observation, we find
$\rho_{matter}(t_{0})\approx \rho_{matter}(t_{RH})[\rho_{\gamma}(t_{0})/
\rho_{\gamma}(t_{RH})]^{3/4}$. We emphasise that this determination of
$\rho_{matter}(t_{0})$ is independent of the unknown dark energy.
Allowing for entropy production, which will increase the photon energy density
relative to the matter energy density, it is usually convenient to 
consider the ratio $n_{matter}/s$ which is equal to
the number of particles of each species
per comoving volume. From the model we can calculate $n_{matter}/s$
at $t_{RH}$, and by particle number conservation the value of this ratio
at the present time $t_0$ is unchanged. Using the present value 
of $s$ (above)
we therefore immediately find $n_{matter}(t_0)$ and from the
mass of the particle type we readily find $\rho_{matter}(t_{0})$, again
independently of the dark energy.

Once we have obtained $\rho_{matter}(t_{0})$, 
from a combination of our model calculation and the
observed CMB temperature, as outlined above, we now use the observed
Hubble constant $H_0$, or equivalently the present day critical
density $\rho_{crit}$, to convert $\rho_{matter}(t_{0})$
into the various $\Omega_{matter}=\rho_{matter}(t_{0})/\rho_{crit}$.
Once $\Omega_{matter}$ is predicted
within some supersymmetric particle physics
based model of inflation, supplemented by measurements of the
CMB temperature and the Hubble constant,
then it is clear that
$\Omega_{DE}$ is also {\em predicted } to be 
\begin{equation}
\Omega_{DE} = 1- \Omega_{matter}
\label{lambda}
\end{equation}
Thus a model of inflation that is capable of predicting
$\Omega_{matter}$ using measurements of the
CMB temperature and the Hubble constant, is also capable of predicting
$\Omega_{DE}$ from Eq.\ref{lambda}.
This sum rule was written down in ref.\cite{sumrule},
including a curvature term and it was discussed there how to 
determine each of the components $\Omega_{DE}$ and $\Omega_{matter}$
from observation. 
What we are saying here is quite different from the empirical approach
to determining the components of this equation discussed
in ref.\cite{sumrule}, and should not be confused with it.
To begin with we are assuming inflation, so that the curvature
contribution is zero. Secondly we are only taking two inputs
from observation, namely the CMB temperature and the value
of the Hubble constant. Given these inputs we have shown how an
inflation model allows us to then {\em calculate} $\Omega_{matter}$,
and hence {\em deduce} $\Omega_{DE}$ from Eq.\ref{lambda}.

At first sight our result appears surprising: how can we
have deduced $\Omega_{DE}$ from a model in which the dark energy
is unspecified? In order to answer this it is useful to compare
two slightly different models of inflation, one in which
the dark energy density is zero and one in which it is non-zero,
but which otherwise predict identical values of 
$\rho_{\gamma}(t_{RH})$ and $\rho_{matter}(t_{RH})$
(in our example this just corresponds to tuning $V_0$ slightly
differently in the two cases leaving all the other parameters
unchanged.) In both cases this will result in the same values of 
$\rho_{matter}(t_{0})$, once $\rho_{\gamma}(t_{0})$ is inputted.
The only difference is that the Friedmann equations, with zero
curvature term due to the inflation assumption, will determine
two different values of the Hubble constant, corresponding to 
two different values of critical density.
The first is given by $\rho_{crit1}=\rho_{matter}(t_{0})$,
and the second involving the sum of two contributions 
$\rho_{crit2}=\rho_{matter}(t_{0})+\rho_{DE}(t_{0})$. 
Since we input the Hubble constant from observation, we know
the true value of $\rho_{crit}$ in our universe, and so we can
discriminate between the two cases from a measurement
of the Hubble constant. More generally, it is clear that,
once the correct supersymmetric particle inflation model is known, and
the present day value of $\rho_{matter}(t_{0})$ is calculated
from it (using the CMB temperature as input), that the present day
value of $\rho_{DE}(t_{0})$ may be deduced from the
Hubble constant $H_0$ which is telling us information about dark energy
by telling us the critical density. From this example it is clear
that our argument gives us no new insight into the cosmic coincidence
question, since a universe without dark energy would simply
correspond to having a different Hubble constant.

Once the importance of the Hubble constant $H_0$ in our
argument is realised, the next question is whether our argument
contains any content at all? The answer must be yes, since
the conclusion relies on non-trivial information coming from
the model, namely the initial condition
for $\rho_{matter}(t_{RH})$ and $\rho_{\gamma}(t_{RH})$
without which it would be impossible to
find $\rho_{matter}(t_{0})$ from the CMB temperature alone,
and without $\rho_{matter}(t_{0})$ it would be impossible to 
deduce $\rho_{DE}(t_{0})$ from the Hubble constant $H_0$.
A related question, is whether our argument is actually useful
in practice, given that at the present time we do not know the
correct model, 
and even if the model were known and all the parameters of that model were
accurately specified, we still would need to know the physics
of preheating and reheating very well.
Also when the argument is generalised to all the forms of matter
and radiation we would need to have a good understanding of the
physics of baryogenesis and a way of calculating the axion
misalignment angle, and so on in order to be able to specify the
present day relic densities of all the component forms of matter.
One could criticise the argument on the grounds that
the accuracy of the deduced density of dark energy is therefore
limited by the accuracy with which the matter density can be
calculated. While this is true, it would seem remarkable
to us to suppose otherwise: while it would be nice to be able
to calculate the DE density much more accurately than the matter
density, this possibility hardly seems very likely.
What our argument gives is a way of calculating
$\Omega_{DE}$ to the same precision
as $\Omega_{matter}$, and this we believe is the best that one can
hope for.

Whether the DE is due to scalar fields, or an incomplete vanishing of
a cosmological constant (corresponding perhaps 
to the universe ending up temporarily in a
vacuum state slightly above a global minimum at zero),
perhaps some of the
parameters that determine it will be related to parameters that also
determine the forms of matter. In the present paper 
we do not specify any particular model for DE
and so we must therefore rely on
observation to determine the equation of state for the DE.
If observation eventually tells us that
$w=-1$ and that the DE is equivalent to a cosmological constant,
then it will be a tremendous challenge to theorists to explain this
(see for example the approach of Bastero-Gil, Mersini, and Kanti
\cite{Mersini:2001su}).
Explaining a vanishing cosmological constant is already proving 
a very difficult question for string theorists, and explaining
a very small one does not apparently make this any easier.
In this case it is possible that the DE question will be around
for a long time. In the meantime progress may be forthcoming 
on determining the supersymmetric
particle inflation model and in determining its
parameters and in being able to use those parameters to calculate
the relic matter densities with increasing accuracy. In such 
a scenario, the one consolation may be that our argument enables
one to then calculate the size of the 
DE density (i.e. the cosmological constant)
even in the absence of any theory of it.

\section{Summary and Conclusion}
We have considered the class of supersymmetric inflation models
in which inflation occurs at the intermediate PQ symmetry breaking
scale. Such models are better motivated than GUT scale inflation
models which face the problems with magnetic monopoles
and domain walls. In intermediate scale supersymmetric inflation
the same theory which is responsible for inflation is
also responsible for the solution to the $\mu$ problem, 
and the strong CP problem. As a consequence
one would generally expect CDM to contain an axion component
in addition to an LSP component. The LSP itself need not be the 
lightest neutralino, but may be a singlino associated with 
the singlet fields which control hybrid inflation and resolve
the $\mu$ problem. The present day relic densities
of the CDM components comprising LSPs and axions may be 
calculated in a given model, using the observed CMB temperature
and Hubble constant, although the axion density will
be subject to the usual uncertainties of the unknown
misalignment angle.

Once right-handed neutrinos are added,
as the recent confirmations of neutrino masses suggests that they
should be, then the possibility of baryogenesis via leptogenesis
seems well motivated, and then the baryon density may be calculated
in a given model. The various relic densities are in principle
calculable in a given model, and are related to each other and to 
other phenomena since the number of parameters involved is
generally smaller than would be the case without a theory.
The same parameters control
cosmology on the one hand and collider physics on the other hand.
For example common soft SUSY breaking
parameters are involved in both inflation and collider physics.
In order to illustrate all these ideas we have described 
an explicit intermediate scale supersymmetric inflation
model which already exists and is quite well studied in the literature 
\cite{Bastero-Gil:1998vn,leptogenesis,sugra}, and many of the general
ideas above are made very explicit in the model, for example
the role of the underlying parameters in determining different
phenomena is demonstrated for this model in Table 1.

Over the next few years there will be considerable 
progress in cosmology from the Map and Planck explorer satellites, and 
in SUSY from the Tevatron and LHC. We believe the time is
ripe for a new closer synthesis of SUSY and inflation, and that
the most promising scenario will involve these theories
meeting at the intermediate scale. We have shown that
in this case one may hope to relate different phenomena
in cosmology and in particle physics in a much closer and
more predictive way than ever before. Finally
we have made the original observation that, given the
value of the CMB temperature and Hubble constant from observation,
an intermediate scale supersymmetric inflation model allows
the present day matter relic density to be calculated,
and hence the present day DE relic density to be determined
from Eq.\ref{lambda} even in the absence of any theory of DE.

\begin{center}
{\bf Acknowledgements}
\end{center}
GK appreciates helpful conversations with S. Carroll, D. Chung,
I. Maor, M. Turner and
L.-T.Wang, and SK is similarly grateful 
to J. Cline, A. Kusenko, A. Liddle and D. Rayner,
and appreciates the support and hospitality of
the Michigan Center for Theoretical Physics 
and PPARC for a Senior Fellowship.

\end{document}